# A Physics-guided Fine-tuned LLM-based Framework for Customized Power Distribution System Feeder Generation

Zhenghao Zhou, *Student Member*, *IEEE*, Yiyan Li, *Member*, *IEEE*, Tao Xu, *Member*, *IEEE*, Yike Guo, Zheng Yan, *Senior Member*, *IEEE*, and Mo-Yuen Chow, *Life Fellow, IEEE*

***Abstract*—Power distribution system feeder models (e.g., IEEE 33-bus system, IEEE 13-bus system, etc.) are cornerstones for conducting power distribution system studies. As real-world feeder models are hard to acquire due to energy security concerns, generating high-quality synthetic feeders becomes an important alternative to satisfy the fast-growing and diversified needs of power system researchers and engineers. In this paper, we propose an LLM-based synthetic feeder generation framework that can achieve end-to-end generation from natural language specifications to physically consistent feeder models. First, Supervised Fine-Tuning (SFT) is performed on a dataset created following physical laws to empower the LLM with syntactic understanding of complex feeder structures. Second, Group Relative Policy Optimization (GRPO) with a specially-designed multi-stage gated reward function is introduced to better align the generation results with user intent and physical constraints. Third, a dual-agent architecture is deployed to refine and evaluate the generated feeders. Specifically, a refinement agent calibrates the feeder model parameters referring to the industrial feeder design standards, while a judge agent provides quality assessments. Case studies demonstrate that the proposed framework generates customizable feeders with valid formats, physical consistency and high engineering applicability.**

## I. Introduction

POWER distribution system test feeders (e.g. IEEE 33-bus system, 13-bus system, 123-bus system, etc.) serve as critical benchmarks for validating smart grid technologies [1], [2]. With the rapid integration of distributed energy resources and the emergence of novel operational paradigms (e.g., virtual power plant, peer-to-peer energy trading, data center co-optimization, etc.), existing standard IEEE test feeders cannot satisfy the increasing and diverse needs of power distribution system researchers and engineers. In such cases, considering that real-world feeders are often hard to access due to energy security concerns, generating synthetic feeders becomes an important alternative and has gained increasing attention in recent years.

Methodologies for synthetic feeder generation can be categorized into two main paradigms: rule-based approaches and machine learning-based approaches. Rule-based methods rely on the empirical analysis of large collections of real feeders to identify and reproduce key statistical metrics and structural properties [3], [4]. These methods typically begin with geographical information and employ computational geometry and graph theory to construct a new feeder topology [5]. Lines are added following certain rules or probabilistic processes to match established empirical laws, such as a specific degree distribution [6] or the heavy-tailed distribution of line impedances [7]. A key focus is placed on ensuring that electrical parameters, like line impedances and transformer ratings, are consistent with the system's voltage level. The validity of the generated feeders is often verified against criteria such as AC power flow convergence [8]-[10]. While these methods are computationally efficient and interpretable, they rely heavily on pre-defined rules and expert knowledge, which may fail to capture the complex, non-linear patterns present in real-world systems and may lead to limited diversity of the generated networks.

Machine learning approaches offer a data-driven solution for synthetic feeder generation, with prominent examples including statistical spatial imitation [11] and deep generative networks [12], [13]. By extracting topological and electrical features directly from existing data, these methods significantly enhance the automation and diversity of the synthetic feeder generation process. Particularly, as a state-of-the-art branch of machine learning, Large Language Models (LLMs) have demonstrated remarkable capabilities in understanding complex data patterns and generating new content. As such, LLMs have been introduced into the power system domain for tasks such as time series analysis [14], scenario generation [15], power flow calculation [16] and fault diagnosis [17], after combining a suite of assistive techniques such as Retrieval-Augmented Generation (RAG) [18], in-context learning [19] and prompt engineering [20] to guarantee the validity and professional quality of the results.

Recently, pioneering efforts have also begun exploring LLMs for power grid and distribution feeder generation [21]-[23]. However, applying large language models to distribution feeder synthesis presents the following challenges that have not been well addressed:

This work was supported by National Natural Science Foundation of China under Grant 52307121. (Corresponding author: Yiyan Li.)

Zhenghao Zhou, Yiyan Li, and Yike Guo are with the College of Smart Energy, Shanghai Non-Carbon Energy Conversion and Utilization Institute, and Key Laboratory of Control of Power Transmission and Conversion, Ministry of Education, Shanghai Jiao Tong University, Shanghai, 200240, China. (e-mail: zhenghao.zhou@sjtu.edu.cn, yiyan.li@sjtu.edu.cn, yike.guo@sjtu.edu.cn).

Tao Xu is with the Key Laboratory of Smart Grid of Ministry of Education, Tianjin University, Tianjin 300072, China (e-mail: taoxu2011@tju.edu.cn).

Zheng Yan is with the Key Laboratory of Control of Power Transmission and Conversion, Ministry of Education, and the Shanghai Non-Carbon Energy Conversion and Utilization Institute, Shanghai Jiao Tong University, Shanghai 200240, China. (e-mail: yanz@sjtu.edu.cn)

Mo-Yuen Chow is with the Global College, Shanghai Jiao Tong University, Shanghai, 200240, China. (email: moyuen.chow@sjtu.edu.cn)

(1) *Long-sequence structural validity.* To adapt feeder synthesis to LLMs, graph-structured feeder data must be converted into textual formats such as JSON. However, a feeder model contains many coupled elements, including buses, branches, source and load nodes, line parameters, and operating constraints, making the serialized text long and highly structured. General-purpose LLMs are prone to errors when generating such long structured outputs, such as missing brackets, invalid keys, inconsistent field types, omitted records, duplicated elements, or mismatched bus and branch identifiers [24], [25]. These format-level errors can make the generated feeder file unparsable or unusable for subsequent power system analysis.

(2) *Physical feasibility*. A structurally valid text sequence may still violate coupled electrical laws or fail AC power flow calculation. The fundamental challenge is that physical rules, such as network connectivity, Kirchhoff laws, and power flow feasibility, are not naturally embedded in the token-level sampling policy of general-purpose LLMs. Existing LLM-based generation studies have begun to use RAG or tool-augmented workflows [21]-[23]. These workflows can improve generation success by introducing external knowledge, solvers, or validation-and-repair loops. However, when the underlying generator is not updated by constraint-derived training signals, physical feasibility remains enforced mainly outside the model rather than being encoded in its token-level generation policy. As a result, feasibility depends on additional inference-time validation or repair, which can increase computational cost and leave the generation process sensitive to the external checker.

(3) *Engineering applicability*. Even if a generated feeder is connected and has convergent AC power flow, it may still be impractical if load allocation, power factor ranges, line impedance, voltage-level matching, or regional planning assumptions are unrealistic. Existing methods often rely on random parameter assignment or simplified assumptions, leaving a gap between calculable feeders and engineering-usable feeders.

(4) *Fair evaluation of generated feeders*. Feeder generation involves syntax correctness, topology validity, physical feasibility, user-intent matching, and engineering realism. Simple pass-or-fail checks are therefore insufficient for comparing different methods. A consistent evaluation mechanism is required to support fair comparison and guide subsequent optimization.

These challenges indicate that LLM-based feeder synthesis should not be treated as a direct prompt-to-file generation problem. A feasible framework must first make feeder models learnable as structured text, then guide the generation policy to satisfy physical and topological constraints, and finally refine and evaluate the generated feeders from an engineering perspective. In other words, the task requires a staged design that separately addresses structured representation learning, physics-aware policy optimization, and engineering-oriented refinement and assessment.

To address these requirements, this paper introduces an LLM-based generative framework for high-fidelity and customized feeder generation. First, Supervised Fine-Tuning (SFT) is used to adapt the model to feeder-specific textual formats and long structured outputs. Second, Group Relative Policy Optimization (GRPO) is adopted to internalize physical and topological constraints into the generation policy through physics-informed rewards. Compared with Proximal Policy Optimization (PPO), GRPO avoids the need for an additional value model, and compared with Direct Preference Optimization (DPO), it does not rely on manually constructed preference pairs [26]-[28]. Third, a dual-agent architecture is developed, where a refinement agent improves engineering applicability and a judge agent provides consistent quality evaluation.

The main contributions are summarized as follows:

1. A procedural SFT dataset generation method is developed to support graph-to-text conversion and long-sequence structural learning. The method uses feeder construction rules to generate structured instruction-output pairs that jointly encode topology, electrical parameters, and operating constraints in JSON text. This provides high-quality training data for teaching the model feeder-specific syntax and reducing malformed structured outputs.
2. A GRPO-based training strategy with a multi-stage gated reward function is designed to embed physical and topological requirements into the generation policy. The reward design considers syntax validity, connectivity, radiality, user-intent alignment, and AC power flow convergence, enabling efficient policy optimization without an additional value model or manually constructed preference pairs.
3. A dual-agent architecture is designed to improve engineering applicability and evaluation fairness. The refinement agent leverages the real-world feeder design standards to iteratively improve the engineering fidelity. The judge agent is developed to provide an objective evaluation for feeder generation quality.

The rest of this paper is organized as follows: In Section II, we introduce the methodology. We demonstrate the case study results in Section III and conclude the paper in Section IV.

## II. Methodology

### A. Problem Formulation

Power distribution system feeders can be represented as graphs, where generators and loads correspond to nodes, and branches constitute the edges. The system's characteristics can then be encoded in feature matrices associated with nodes and edges. However, LLMs are trained predominantly on vast corpora of textual data and are inherently designed for understanding and generating natural language. Consequently, representing feeder systems through structured text rather than raw numerical or graphical formats is a more suitable approach for LLMs. As shown in Fig. 1, textual representations of power systems can be readily converted into JSON format, which aligns well with the structured output requirements of LLMs. With reference to the MATPOWER format [29], we define a JSON-formatted feeder system $f$ as follows:

$$f = \left\{ k_i : \mathbf{m}^i \mid k_i \in \{baseMVA, bus, gen, branch\} \right\} \quad (1)$$

where $k_i$ is the $i^{th}$ key, and $\mathbf{m}^i$ represents the corresponding physical information.

The generation process is fundamentally a text-to-text transformation. Given a customized feeder requirement provided as a natural language prompt, the large language model generates a textual description of a feasible feeder system, encoded in structured format.

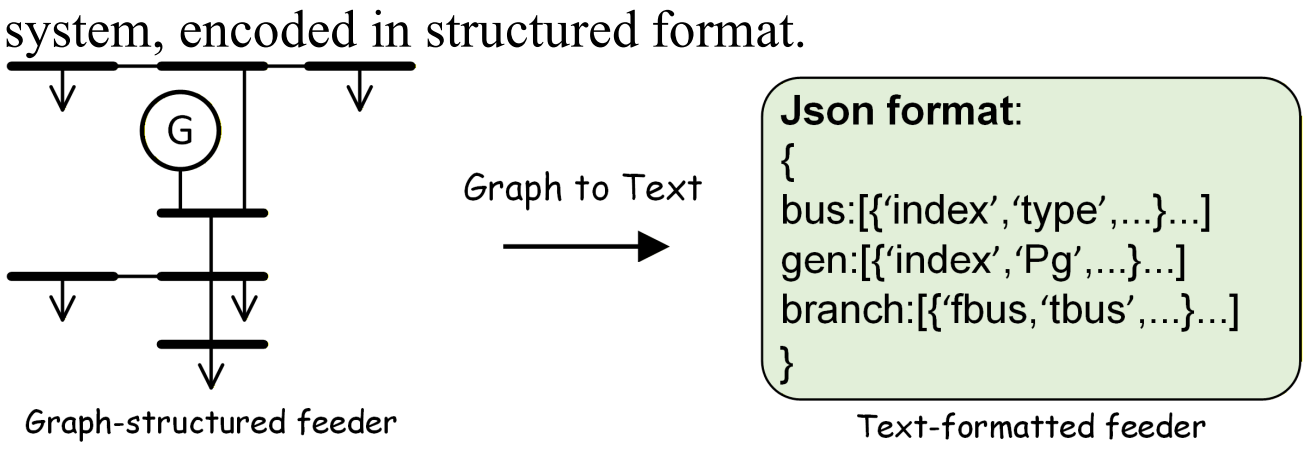

Fig. 1. The definition of graph-to-text conversion.

## B. Framework Overview

Open-source LLMs exhibit strong capabilities in general natural language understanding and generation. However, they often lack sufficient domain-specific knowledge to produce accurate or contextually appropriate outputs in specialized fields like power systems. Parameter-efficient fine-tuning on domain-specific datasets is an effective method to bridge this gap and enhance the expertise in targeted applications.

In this work, we adapt a training strategy to specialize the model for feeder system generation. First, we perform SFT (Section II.D) using a high-quality dataset (Section II.C) of paired input queries and corresponding output feeders. This stage enables the model to acquire fundamental task comprehension and generate structurally valid outputs, establishing a reliable initial policy $\pi_{\text{SFT}}$. Second, we apply GRPO (Section II.E) to further improve the generation policy. Based on SFT, this RL phase directly optimizes the model based on reward signals derived from the quality of the generated feeder systems. Subsequently, we design a dual-agent architecture (Section II.F) to refine and evaluate the generated feeders. The refinement agent enhances engineering applicability by incorporating standards, followed by the judge agent which delivers a quantifiable quality assessment.

In the following sections, we detail the four main components of our methodology: fine-tuning dataset synthesis, SFT, GRPO, and the dual-agent architecture.

## C. Procedural Dataset Synthesis for Fine-Tuning

To address the lack of large-scale high-fidelity datasets, a computationally efficient framework is proposed to generate feeder samples. As shown in Fig. 2, this framework simulates a network construction process where the topology and parameters are synthesized through integrated calculations based on electrical principles.

### 1) Initialization

The generation process begins by defining an initial prompt $q$ that combines the grid type, spatial distribution, topology, and quantities of load nodes and generators. Based on this prompt, system boundary conditions are initialized within a two-dimensional space. A finite set of buses is partitioned into a slack bus, active generator buses, and demand buses. Each node is assigned spatial coordinates and initial power demands. Distinct grid types are differentiated through parameter sampling. Rural feeders utilize low density random placement and low load powers, while urban and industrial networks utilize high density spatial distributions and elevated load levels. This sampling strategy ensures that diverse spatial layouts and load characteristics are captured.

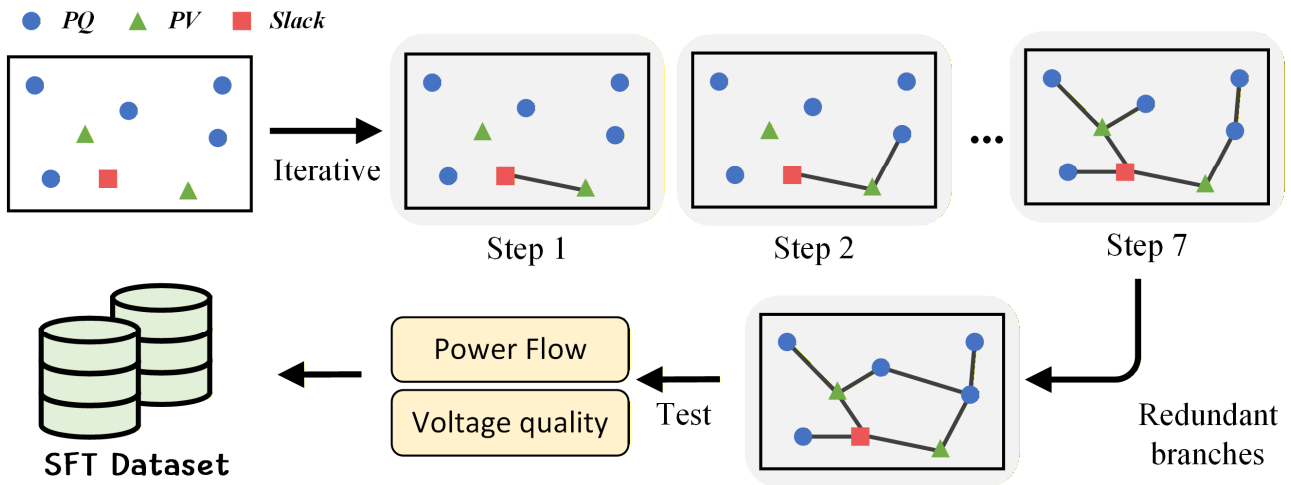

Fig. 2. The workflow of dataset synthesis for fine-tuning.

### 2) Iterative Synthesis of Topology and Parameters

The core framework constructs the network through an iterative algorithm that assigns physical branch parameters while ensuring compliance with operational constraints. The algorithm maintains a connected node set $\mathcal{C}$ and an unconnected set $\mathcal{U}$. In each iteration, a target node $j^*$ is selected from $\mathcal{U}$ based on a heuristic electrical stress metric $S_j$. This metric prioritizes nodes exhibiting high power demand and proximity to existing infrastructure:

$$S_j = |P_j| \cdot \min_{i \in \mathcal{C}} \| p_i - p_{j^*} \|^2 \tag{2}$$

where $p_i$ and $p_{j^*}$ denote the spatial coordinates of nodes $i$ and $j^*$, and $P_j$ represents the net active power at the corresponding node.

Following node selection, a physics-guided parameter assignment procedure is executed. All feasible connections from source nodes $i \in \mathcal{C}$, to $j^*$ are evaluated. To guarantee physical feasibility, an inverse design approach determines the valid resistance and reactance values. Using the LinDistFlow approximation [30] for radial networks, the voltage drop is linearized as:

$$V_i - V_{j^*} \approx (P_{j^*} R_{ij} + Q_{j^*} X_{ij}) / V_{\text{base}} \tag{3}$$

where $P_{j^*}$ and $Q_{j^*}$ denote the net active and reactive power at node $j^*$, and $V_{\text{base}}$ is the nominal base voltage of the primary distribution feeder. Assuming a fixed reactance to resistance ratio $\kappa$ = $X_{ij}$ / $R_{ij}$, the maximum allowable resistance that satisfies a prescribed minimum voltage threshold $V_{\min}$ is derived as:

$$R_{\max} = (V_{\text{base}} (V_i - V_{\min})) / (P_{j^*} + \kappa Q_{j^*}) \tag{4}$$

The maximum permissible resistance per unit length is computed by dividing $R_{\max}$ by the Euclidean length of the branch. The final connection is established by selecting the source node $i$ that provides mathematically feasible physical parameters within these strict limits. The selected branch is incorporated into the connected set $\mathcal{C}$, and the process repeats until all nodes are connected.

### 3) Redundant Connection

To establish effective tie lines rather than redundant local loops, a topology aware connection strategy is implemented. Candidate branches are identified from unconnected node pairs that provide genuine structural redundancy. The topological distance $d_{topo}(i, j)$ within the radial tree between any candidate node pair $(i, j)$ is required to exceed a predefined threshold $d_{th}$. This threshold is dynamically determined based on the maximum depth $D_{max}$ of the initial radial backbone. Valid candidate pairs are sorted by their Euclidean distance. The shortest branches are selected based on the predefined redundancy requirements and incorporated into the grid topology to form a meshed network.

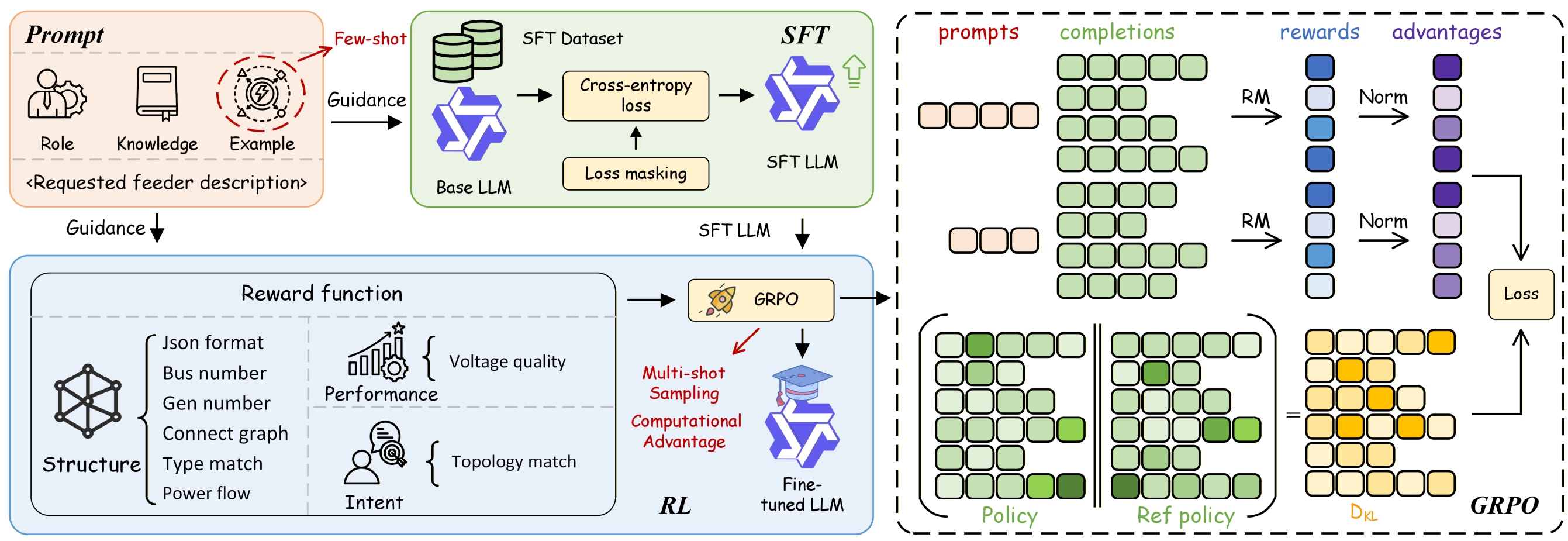

Fig. 3. The proposed training framework. Stage 1 (SFT): A base LLM is fine-tuned on a dataset of expert demonstrations using cross-entropy loss to adapt to the target syntax and domain knowledge. Stage 2 (RL with GRPO): The SFT model undergoes further optimization via GRPO. For each prompt, the policy generates multiple completions, which are scored by a multi-dimensional reward function evaluating structural correctness, performance metrics, and intent alignment. These rewards are normalized within each sampled group to form group-relative advantages, guiding the model to maximize expected return while maintaining training stability. This process leverages computational advantage through parallel sampling and efficient gradient updates to refine complex reasoning capabilities.

#### 4) Validation and Dataset Curation

To ensure physical consistency, standard iterative methods are utilized to solve the full nonlinear alternating current power flow equations. This solution process dictates that all buses satisfy Kirchhoff's laws and the corresponding power balance equations:

$$P_i = V_i \sum_{j=1}^{n} V_j (G_{ij} \cos\theta_{ij} + B_{ij} \sin\theta_{ij}) \quad (5)$$

$$Q_i = V_i \sum_{j=1}^{n} V_j (G_{ij} \sin\theta_{ij} - B_{ij} \cos\theta_{ij}) \quad (6)$$

where $V_i$ represents the voltage magnitude, $\theta_{ij}$ is the phase angle difference, and $G_{ij}$ and $B_{ij}$ are the conductance and susceptance components of the nodal admittance matrix. To validate the physical feasibility of the generated networks, two strict criteria are enforced. First, the power flow calculation must achieve mathematical convergence. Second, the voltage magnitude $V_i$ at each bus must remain within standard operational limits, defined between 0.95 and 1.05 per unit. Only samples satisfying both requirements are retained in the final synthetic dataset, while divergent or voltage-violating cases are directly discarded. The SFT dataset is defined as $\mathcal{D}=\{(q_i, y_i)\}_{i=1}^{N}$, where $q_i$ represents a user query and $y_i$ denotes the textual representation of the corresponding feeder system.

### D. Supervised Fine-Tuning

As shown in Fig. 3, SFT is the foundational stage for adapting a pre-trained language model to a specific downstream task and its expected response format.

#### 1) Prompt Engineering

To effectively guide the generation process, a highly structured prompt template is designed. The prompt comprises a static system instruction and a dynamic user query. The system instruction restricts the output exclusively to valid text formats. It explicitly defines the column dimensions for bus, generator, and branch matrices. Furthermore, fundamental electrical rules are embedded into the instruction to promote initial physical feasibility. For instance, the prompt mandates the presence of a single slack bus and enforces strict mapping between generator nodes and their corresponding parameter arrays. A feeder example is also integrated to provide a few-shot demonstration. Following the system instruction, the dynamic user query provides the specific design targets. This query includes the grid category, topological requirements, and exact quantities of specific nodes. This structured approach ensures that the model precisely translates textual intentions into standardized power system data structures.

#### 2) Training Process

This process leverages a synthesized high-quality dataset $\mathcal{D}=\{(q_i, y_i)\}_{i=1}^{N}$. The objective is to optimize the model parameters $\theta$ to maximize the conditional probability $P_\theta(y \mid q)$ of generating the target feeder $y$ given the prompt $q$. The generation task is formulated as standard autoregressive sequence modeling. The model minimizes the cross-entropy loss over the target sequence tokens of length $T$:

$$\mathcal{L}_{SFT} = -\frac{1}{T} \sum_{t=1}^{T} \log P_\theta (y_t \mid q, y_{<t}) \quad (7)$$

During implementation, the prompt $q$ and the target response $y$ are concatenated into a single continuous sequence. In standard autoregressive modeling, the model predicts the next token at every position across the entire sequence. However, the objective of instruction tuning is exclusively to train the model to generate the response $y$ conditioned on $q$. The model does not need to learn the text distribution of the user prompts. To address this, a loss masking strategy is applied. The loss calculation for the tokens belonging to the prompt $q$ is masked to zero. The cross-entropy loss is computed and backpropagated exclusively on the token positions corresponding to the target response $y$. By minimizing this loss, the SFT model $\pi_{SFT}$ is established. This model acquires basic structural generation capabilities and serves as the initial reference policy for the subsequent RL optimization stage.

### E. Group Relative Policy Optimization

To optimize the model's performance on specific tasks, this work employs the GRPO algorithm to fine-tune the pre-trained model via RL. GRPO is an efficient variant of PPO, specifically designed for the alignment of LLMs. In the traditional PPO-Critic framework, a separate value network (i.e., the Critic) must be trained to estimate the expected state returns, which introduces substantial memory and computational overhead in LLM scenarios. GRPO circumvents this issue through a novel advantage estimation method.

As shown in Fig. 3, for each prompt $q$, the policy $\pi_\theta$ generates a group of responses $\{o_1, \ldots, o_G\}$, which are evaluated by a reward model $R_\varphi$ to obtain scalar rewards $r_g$. A group-relative advantage is then constructed by normalizing rewards within each sampled group. The GRPO objective is formulated as a token-level policy optimization problem, where updates are performed with respect to the behavior policy used during sampling. The optimization objective and advantage definition are given as follows:

$$\mathcal{J}(\theta) = \mathbb{E}\left[\frac{1}{G}\sum_{g=1}^{G}\frac{1}{T_g}\sum_{t=1}^{T_g}\rho_\theta(o_{g,t})A_g - \beta \mathrm{KL}(\pi_\theta \| \pi_{\mathrm{ref}})\right] \quad (8)$$

$$A_g = \frac{r_g - \mu_R}{\sigma_R + \epsilon} \quad (9)$$

where $\rho_\theta(o_{g,t})$ denotes the importance ratio between the current policy and the behavior policy, $A_g$ denotes the group-relative advantage of response $o_g$, and $\mu_R$, $\sigma_R$ are the mean and standard deviation of rewards within the sampled group. The KL divergence regularizes the policy $\pi_\theta$ towards a fixed reference model $\pi_{\mathrm{ref}}$, ensuring training stability and preventing excessive deviation during optimization.

To provide an effective learning signal, a multi-stage additive reward function $R(q, o)$ is adopted, which evaluates physical feasibility, semantic consistency, and task-level performance. The total reward $R_{\mathrm{total}}$ is formally defined as the sum of three components:

$$R_{\mathrm{total}} = w_1 R_{\mathrm{struct}} + w_2 (R_{\mathrm{volt}} \cdot R_{\mathrm{inst}}) - P_{\mathrm{cheat}} \quad (10)$$

where $R_{\mathrm{struct}}$ represents structural and physical feasibility, $R_{\mathrm{intent}}$ denotes intent alignment, and $R_{\mathrm{volt}}$ is the voltage quality metric acting as a gating factor, and $P_{\mathrm{cheat}}$ is a severe penalty for reward hacking.

The reward calculation is governed by a strict checkpoint mechanism. The model's output must first pass all validations within the foundational $R_{\mathrm{struct}}$ phase. Failure at any step in this stage triggers an early exit mechanism: subsequent evaluations are bypassed, the intent reward defaults to zero, and the model receives only the partial baseline score accumulated thus far. Crucially, if the model successfully converges, the mixed gating mechanism explicitly scales the intent reward $R_{\mathrm{intent}}$ by the physical voltage quality $R_{\mathrm{volt}} \in [0,1]$. This ensures that the reward for topological customization is heavily attenuated if the underlying grid suffers from severe voltage violations, thereby encouraging optimization on outputs that have passed the physical feasibility checks.

The component $R_{\mathrm{struct}}$ comprises a sequential validation cascade. It commences with syntactic validation to ensure the output is JSON with the correct matrix dimensions. It then proceeds to electrical rule validation, which mandates the absolute uniqueness of the slack bus and strictly prohibits pure load buses from appearing in the generator matrix. Concurrently, an anti-cheat assessment evaluates the total active load; if it falls below a minimum physical threshold, $P_{\mathrm{cheat}}$ is triggered to penalize the generation of empty grids. Subsequently, a breadth-first search algorithm verifies graph connectivity, and a non-linear AC power flow solver determines convergence, acting as the ultimate gatekeeper for physical validity.

The intent reward $R_{\mathrm{intent}}$ is computed only if all $R_{\mathrm{struct}}$ validations are successfully passed. To provide continuous gradients and avoid local optima, the topology alignment employs smooth exponential and Gaussian kernels. To universally penalize unrealistic super-hub nodes, an exponential decay $P_{\mathrm{deg}}$ based on the maximum node degree $D_{\mathrm{max}}$ is applied:

$$P_{deg} = 1/(1 + \mathrm{e}^{\lambda \cdot (D_{max} - \theta_{hub})}) \quad (11)$$

where $\theta_{hub}$ is the maximum allowable degree threshold (e.g., $\theta_{hub} = 8$).

The specific topological rewards adapt to the requested grid type. For a radial grid, the reward function continuously penalizes the cyclomatic number $c$ and aligns the leaf node ratio $\rho_{leaf}$ to an engineering target $\theta_{leaf}$:

$$R_{\mathrm{rural_topo}} = (\alpha\, \mathrm{e}^{-k_1 c^2} + \beta\, \mathrm{e}^{-k_2(\rho_{leaf} - \theta_{leaf})^2}) \cdot P_{deg} \quad (12)$$

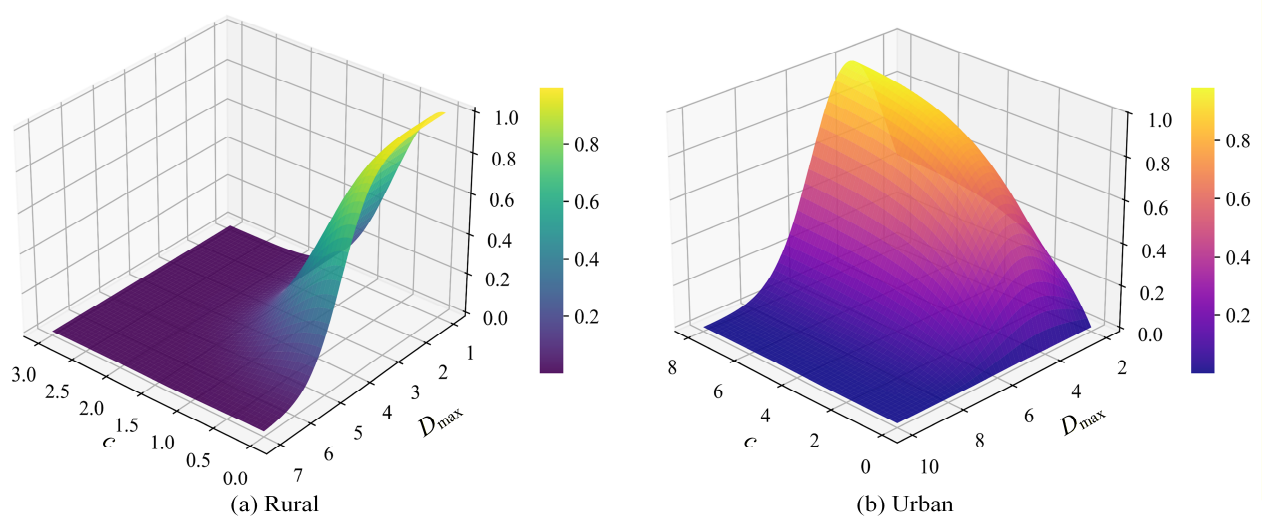

Fig. 4. Visualization of the continuous topological intent reward landscapes across different grid customization scenarios: (a) Rural radial topology and (b) Urban mesh topology.

Conversely, for a mesh-like urban grid, the reward shifts the global optimum to reflect practical interconnected infrastructures. It employs an asymptotic growth function to encourage loop formation ($c$>0) and utilizes a Gaussian kernel to anchor the maximum node degree at an ideal intersection hub value (e.g., $D_{\mathrm{max}} \approx 4$), rather than uniformly penalizing all degree increments:

$$R_{\mathrm{urban_topo}} = (\gamma(1 - \mathrm{e}^{-k_3 c}) + \delta\, \mathrm{e}^{-k_4(D_{max} - 4)^2}) \cdot P_{deg} \quad (13)$$

As shown in Fig. 4(a), the rural grid optimum is strictly anchored at $c$=0 with a smooth penalty for high-degree super-hubs. Conversely, Fig. 4(b) shifts the urban grid optimum to encourage interconnectivity $c$>0 and anchors the node degree around $D_{\mathrm{max}} \approx 4$, reflecting ideal mesh geometry. Both surfaces feature entirely smooth gradients without non-differentiable kinks, preventing vanishing gradients and ensuring stable RL convergence.

### *F. Dual-Agent Refinement and Judge Framework*

Following the GRPO alignment phase, the policy model $\pi_\theta$ demonstrates the capability to generate structurally valid topological data. However, due to inherent limitations in understanding specialized power systems engineering, the initially generated feeder system $f$ may still exhibit parameter homogenization or deviate from strict industrial constraints. To address this, a post-processing framework utilizing LLMs via a dual-agent architecture is introduced to further align the outputs with complex engineering practices. The dual-agent architecture is shown in Fig. 5.

#### *1) Physics-Informed Refinement Agent*

The physics-informed refiner agent focuses on rectifying engineering rule violations and ensuring the physical feasibility of the initially generated feeder $f^{(0)}$. A set of practical power distribution design standards (e.g., DL/T 5729, DL/T 5220, GB 50052) is formalized into a textual rule set $\mathcal{R}$. This rule set is injected into the agent's system prompt to provide prior domain knowledge regarding specific grid categories, such as distinct load densities and typical impedance ratios.

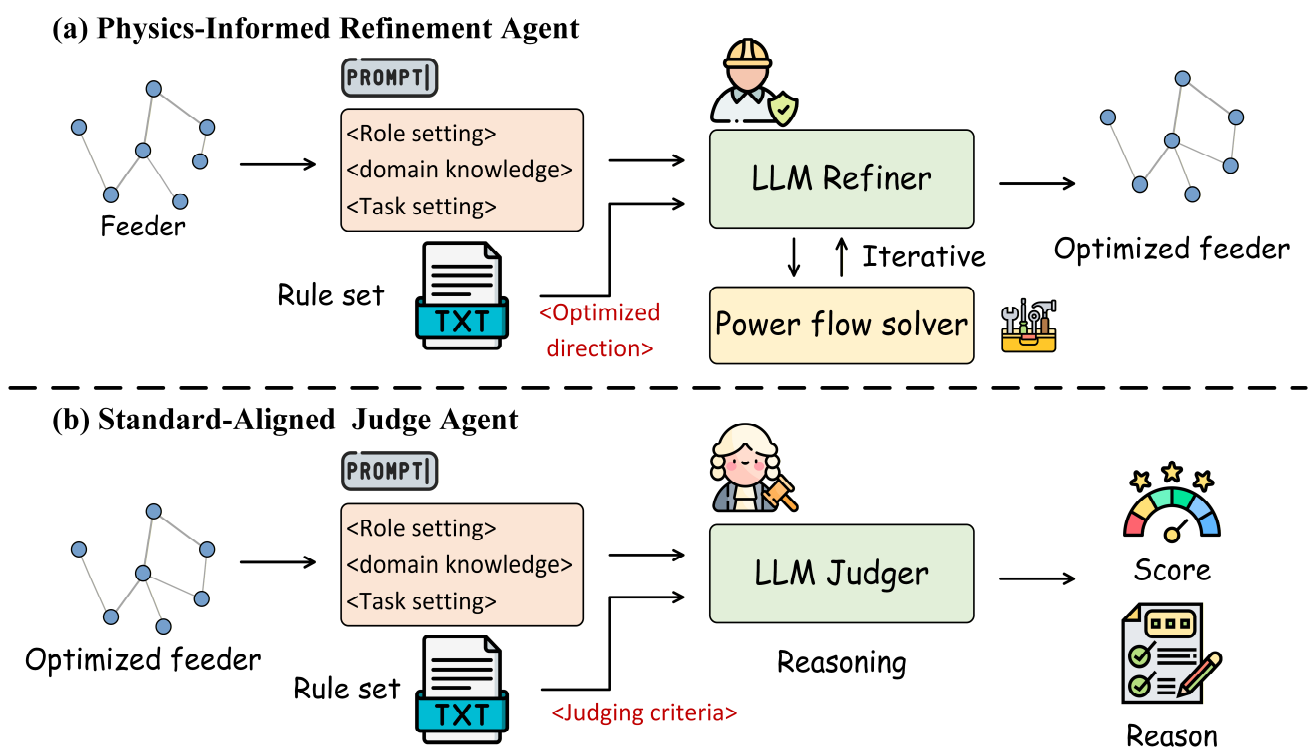

Fig. 5. The proposed dual-agent architecture. (a) The refinement agent iteratively interacts with a power flow solver to optimize the initial feeder based on engineering rules. (b) The judge agent acts as an expert evaluator, scoring the engineering quality of the optimized feeder and providing objective reasoning.

Furthermore, to guarantee the convergence of non-linear AC power flow equations, the refiner agent interacts with a local power flow solver, which acts as an external validation environment $\mathcal{E}$. The refinement process operates iteratively. At each step $t$, the environment evaluates the current feeder and returns a set of physical or engineering errors $e^{(t)} = \mathcal{E}(f^{(t)})$. The agent then updates the feeder parameters conditioned on the initial prompt $q$, the rule set $\mathcal{R}$, and the feedback $e^{(t)}$:

$$f^{(t+1)} = \text{Agent}_{refine}(f^{(t)} | \; q, \mathcal{R}, e^{(t)}) \tag{14}$$

This closed-loop tool-calling mechanism continues until no violations are detected ($e^{(t)} = \varnothing$) or a predefined maximum iteration limit is reached, yielding the optimized feeder $f_{opt}$.

*2) Standard-Aligned Judge Agent*

To objectively evaluate the engineering quality of the generated feeder systems, we introduce the Judge Agent Score (JAS). JAS serves as a comprehensive metric designed to emulate domain experts' judgments regarding the physical fidelity and standard alignment of the generated grids. For a single generated feeder $f$ conditioned on user query $q$, the score is formally defined as:

$$\text{JAS}(f, q) = \text{Agent}_{judge}(f | \; q, \mathcal{R}) \tag{15}$$

where $\mathcal{R}$ represents the injected rule set of engineering standards, and the output is a scalar bounded in [0, 10].

Furthermore, to evaluate the overall quality of a generated dataset containing $n$ samples, we calculate the Mean Judge Agent Score (MJAS):

$$\text{MJAS} = \frac{1}{n}\sum_{k=1}^{n} \text{JAS}(f^{(k)}, q^{(k)}) \tag{16}$$

## III. Case Study

### *A. Experimental Setup and Metrics*

The experiments employ a two-stage training pipeline comprising SFT and GRPO. All models are trained on NVIDIA A100 GPUs. The base model is Qwen3-8B. The training utilizes 4-bit quantization and Low-Rank Adaptation to optimize memory usage. The adapters are applied to the attention and feed-forward modules with a rank of 16, a scaling factor of 32, and a dropout rate of 0.05.

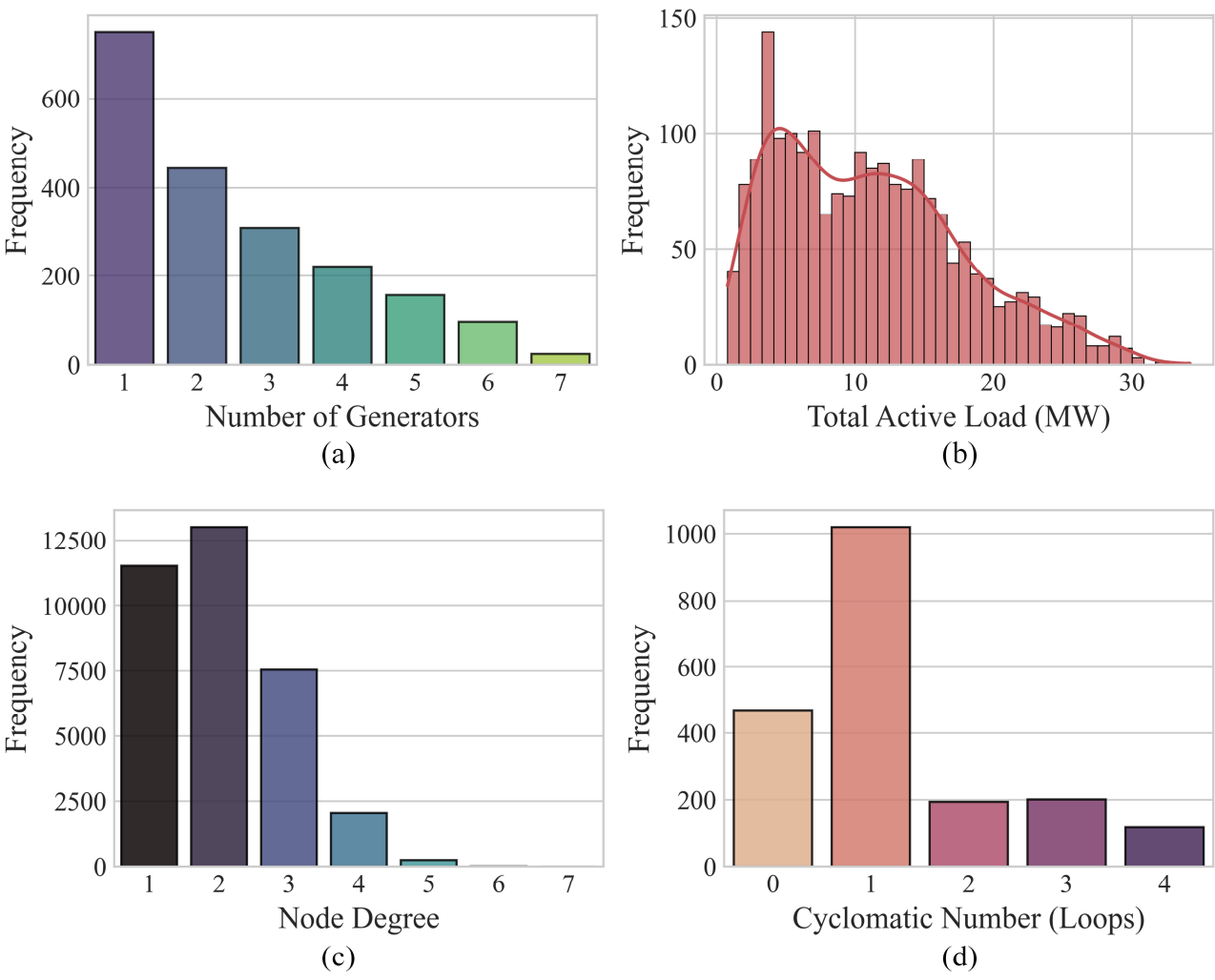

Fig. 6. Statistical distributions of the SFT dataset. (a) shows the frequency of the number of generators per feeder system. (b) presents the distribution of the total active load. (c) displays the node degree distribution across all generated grids. (d) illustrates the distribution of the cyclomatic number to quantify the presence of redundant loops within the network structures.

In the SFT stage, the model is trained for 5 epochs using an 8-bit AdamW optimizer. The learning rate is set to $2 \times 10^{-5}$ with a weight decay of 0.01. The batch size is 4, gradient accumulation steps are 8, and the linear scheduler includes 50 warmup steps. The dataset is preprocessed to remove thought process tags, and loss masking is applied to prompt tokens to exclude them from gradient computation. In the GRPO stage, the model is trained for 2 epochs using a Paged AdamW 8-bit optimizer. The learning rate is $5 \times 10^{-6}$ with a cosine scheduler. The framework samples 4 generation trajectories per prompt to compute relative advantages. The Kullback-Leibler divergence penalty coefficient is set to 0.04. The generation temperature is 0.85 and the top-p value is 0.9 to ensure sufficient exploration. Our implementation integrates Unsloth to accelerate fine-tuning and deploys vLLM to achieve low-latency inference.

Following the GRPO, the post-processing framework utilizes the Qwen3.6-Max-Preview model to construct the dual-agent system. The physics-informed refinement agent and the standard-aligned judge agent are governed by a formalized rule set derived from engineering standards, ensuring the physical fidelity and industrial compliance of the generated grids.

The proposed training paradigm utilizes specific datasets for both stages. For the first stage, the procedural generation algorithm creates a comprehensive dataset. This dataset contains 2000 pairs of text prompts and feeder systems. Fig. 6 presents the statistical distributions of the SFT dataset. Each sample links a descriptive text prompt with a feeder system in the text format. These generated systems feature diverse topologies and guarantee the convergence of power flow. Statistical distributions confirm the physical realism and structural diversity of the data. The number of generators per system ranges from 1 to 7. The total active load spans a broad spectrum up to approximately 35 megawatts. Topological complexity is further demonstrated through node connectivity metrics. A high frequency of unit node degrees indicates abundant terminal load points, while the presence of higher degrees reflects complex intersection hubs. Furthermore, the cyclomatic number distribution covers both strictly radial configurations with zero loops and highly interconnected mesh networks with multiple loops. This collection serves as the foundation to align the model with the syntax of the domain and basic physical structures. For the subsequent RL stage, a separate dataset provides 200 manually crafted text descriptions of feeders. Unlike the SFT dataset, this collection solely provides the prompts without reference responses. These prompts trigger the exploratory generation of the model. This process refines the policy through a reward mechanism with multiple stages.

To assess the quality of the generated feeders, we design a comprehensive evaluation framework. This framework evaluates the generated outputs from structural, physical, and semantic perspectives. Let $N$ denote the total number of generated samples $\mathcal{O}$.

Format Pass Rate (FPR): It measures the proportion of outputs that form valid JSON objects and contain the correct matrix dimensions. Let $\mathbb{I}_{format}$ be an indicator function that equals 1 if the output is structurally valid and 0 otherwise.

$$FPR = \frac{1}{N}\sum_{o\in\mathcal{O}} \mathbb{I}_{format}(o) \tag{17}$$

Physical Convergence Rate (PCR): It is computed on the structurally valid samples (denoted as the subset $\mathcal{O}_{valid} \subseteq \mathcal{O}$, where $N_{valid}$=|$\mathcal{O}_{valid}$|). This rate represents the percentage of grids that successfully solve the non-linear alternating current power flow equations. Let $\mathbb{I}_{conv}$ equals 1 if the grid converges and 0 otherwise.

$$PCR = \frac{1}{N_{valid}}\sum_{o\in\mathcal{O}_{valid}} \mathbb{I}_{conv}(o) \tag{18}$$

Voltage Qualification Rate (VQR): It evaluates the operational performance of the converged grids (denoted as the subset $\mathcal{O}_{conv} \subseteq \mathcal{O}_{valid}$, where $N_{conv}$=|$\mathcal{O}_{conv}$|) by calculating the proportion of secure nodes. Let $B_o$ be the total number of buses in the converged grid $o$, and $\mathbb{I}_{volt}$ equal 1 if the voltage $V_{o,v}$ falls within the statutory limits (e.g., $0.95<V_{o,v}<1.05$p.u.) and 0 otherwise.

$$VQR = \frac{1}{N_{conv}}\sum_{o\in\mathcal{O}_{conv}} \left(\frac{1}{B_o}\sum_{v=1}^{B_o} \mathbb{I}_{volt}(V_{o,v})\right) \tag{19}$$

Generator Constraint Rate (GCR): The ratio of converged feeders where the active and reactive power generation of all units satisfy capacity limits. Let $\mathbb{I}_{gen}$ equal 1 if all generators in grid $o$ respect capacity bounds (i.e., $P_{min} \le P_g \le P_{max}$ and $Q_{min} \le Q_g \le Q_{max}$) and 0 otherwise.

$$GCR = \frac{1}{N_{conv}}\sum_{o\in\mathcal{O}_{conv}} \mathbb{I}_{gen}(o) \tag{20}$$

Line Constraint Rate (LCR): The proportion of generated models where the apparent power flow of distribution lines is constrained to the thermal limit. Let $\mathbb{I}_{line}$ equal one if apparent power flow $S_l$ across every branch $l$ in grid $o$ satisfies $S_l \le S_l^{max}$, and 0 otherwise.

$$LCR = \frac{1}{N_{conv}}\sum_{o\in\mathcal{O}_{conv}} \mathbb{I}_{line}(o) \tag{21}$$

Full Constraint Rate (FCR): It represents the proportion of structurally generated models that strictly satisfy all aforementioned constraints.

$$FCR = \frac{1}{N}\sum_{o\in\mathcal{O}} \left(\mathbb{I}_{format}\cdot\mathbb{I}_{conv}\cdot\mathbb{I}_{volt}\cdot\mathbb{I}_{gen}\cdot\mathbb{I}_{line}\right) \tag{22}$$

Average Reward Score (ARS): The mean scalar reward assigned by the reward model across all test samples $\mathcal{O}$.

### B. Baseline Comparison

To comprehensively evaluate the proposed method, we select four representative baseline models covering various learning paradigms. The first baseline is the zero-shot large language model, which assesses the innate domain understanding capability. The second is the few-shot in-context learning approach, evaluating the ability to follow topological and physical rules through prompt observations. The third baseline is the SFT model, demonstrating the performance of traditional data-driven alignment. The fourth baseline is the DPO model, representing mainstream contrastive preference learning methods. These models are compared against our proposed GRPO method to highlight the necessity of embedding strict physical gating mechanisms during the training phase.

TABLE I
MODEL PERFORMANCES

| Model | Zero-shot LLM | Few-shot LLM | SFT | SFT+DPO | SFT+GRPO |
|---|---|---|---|---|---|
| FPR | 0 | 0.775 | 0.990 | 0.985 | 1 |
| PCR | 0 | 0.384 | 0.950 | 0.985 | 0.980 |
| VQR | 0 | 0.263 | 0.825 | 0.910 | 0.935 |
| GCR | 0 | 0.247 | 0.830 | 0.840 | 0.957 |
| LCR | 0 | 0.384 | 0.990 | 0.989 | 0.990 |
| FCR | 0 | 0.179 | 0.710 | 0.879 | 0.895 |
| ARS | 0 | 0.418 | 0.849 | 0.855 | 0.889 |

As summarized in Table I, a comprehensive quantitative evaluation of the generative models across various training methods is provided. The zero-shot LLM fails completely to generate valid feeder configurations and records scores of zero across all metrics. The inclusion of in-context examples via the few-shot LLM approach yields marginal improvements, yet the performance remains fundamentally inadequate for practical physical simulations. The application of supervised fine-tuning establishes the critical foundation for domain adaptation. The SFT model demonstrates a substantial performance improvement compared to the baseline models. The FPR reaches 0.990, and the PCR achieves 0.950. However, the FCR of 0.710 highlights the limitations of purely data-driven imitation. The SFT model struggles to guarantee absolute physical feasibility under strict and simultaneous operational constraints. The subsequent reinforcement learning phase effectively resolves these physical limitations. While the SFT+DPO model improves the overall FCR to 0.879 and increases the GCR to 0.840 compared to the SFT baseline, it still struggles to balance all operational limits simultaneously. This performance ceiling exists because DPO relies on relative preference datasets, making it inadequate for enforcing absolute physical boundaries, such as strict active and reactive power limits. The SFT+GRPO framework achieves the optimal performance balance. This architecture achieves a 100% FPR and records the highest FCR of 0.895. Furthermore, the GRPO alignment successfully internalizes complex equipment limitations, elevating the GCR to a peak value of 0.957 and securing the highest average reward score of 0.889. This progression validates the effectiveness of the training paradigm: the SFT stage elevates the fundamental generative capabilities of the model, while the subsequent GRPO stage aligns the generation policy with human customization preferences and strict operational constraints.

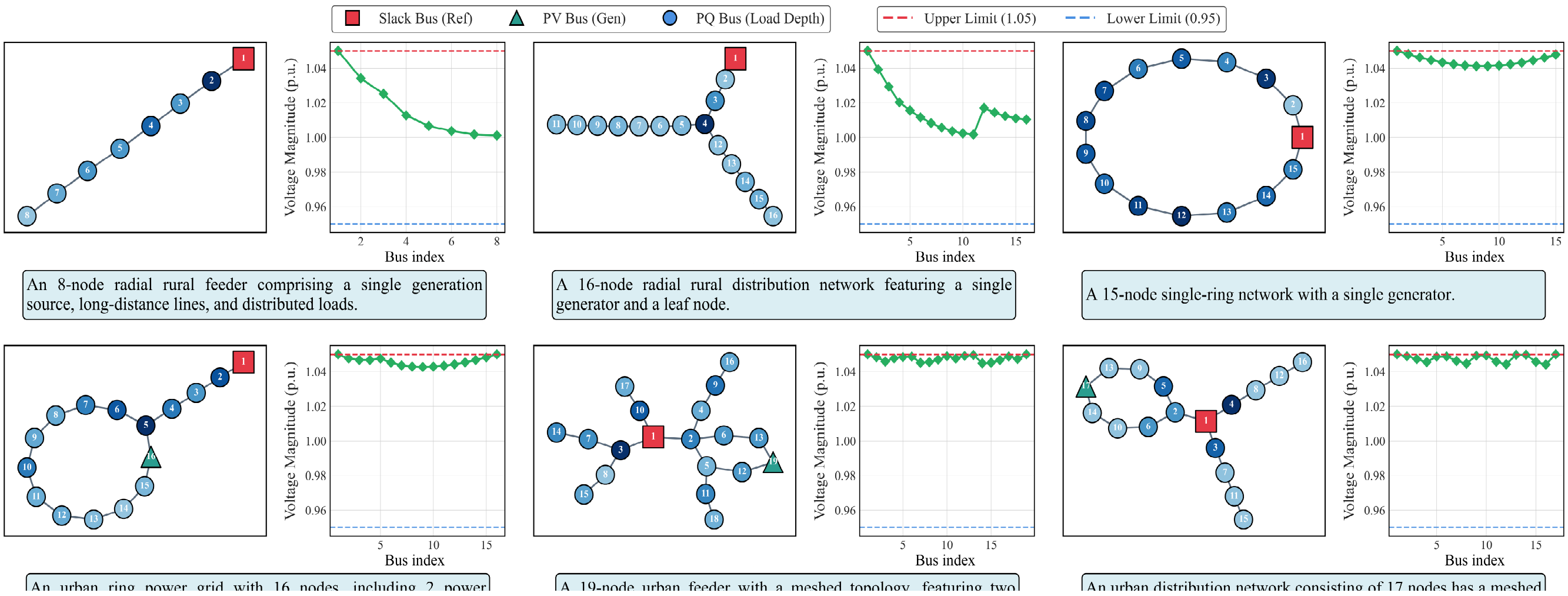

Fig. 7. Visual analysis of samples. Left: The network topology where node shapes represent bus types (Slack, PV, and PQ). The colormap of the PQ buses (blue circles) indicates varying active power load depths, with darker shades representing heavier localized loads. Right: The corresponding AC power flow voltage profile. The model demonstrates implicit physical reasoning by strategically dispatching generator voltages (PV and Slack buses) near the 1.05 p.u. upper limit to proactively counteract downstream voltage drops, ensuring all nodes remain within the specified [0.95, 1.05] p.u. operational limits.

The visual analysis of the generated samples demonstrates a high degree of semantic alignment between the input text and the output network. As shown in Fig. 7, prompts requesting radial rural feeders produce distinct tree structures. Conversely, prompts specifying urban ring or meshed topologies successfully yield complex interconnected networks. The spatial layout and the allocation of load or generation nodes adhere to the customized requirements.

The corresponding voltage profiles confirm the physical realism of the generated systems. All node voltages are constrained within the specified operational limits of 0.95 to 1.05 per unit. The voltage curves exhibit natural physical behaviors that correspond to their specific topologies. Radial networks with a single generator display a clear and gradual voltage drop along the feeder lines. In contrast, urban meshed grids with multiple power generation nodes maintain a significantly flatter and more robust voltage profile. The policy optimization successfully learns to dispatch the generator voltages near the upper limit to proactively counteract downstream voltage decay. In the tested cases, this behavior is associated with successful power flow convergence and improved voltage quality across different customized feeder structures.

## C. Engineering Improvement

The engineering value of the generated distribution networks is evaluated using a judge agent powered by the Qwen3.6-Max-Preview model. The evaluation results indicate that the data generated by the GRPO model achieves an MJAS of 4.19. Following the intervention of the refinement agent, the MJAS increases to 8.05. This improvement confirms the effectiveness of the dual-agent architecture in aligning generated feeders with practical engineering requirements. The refinement agent achieves this alignment through parameter calibration and topological restructuring.

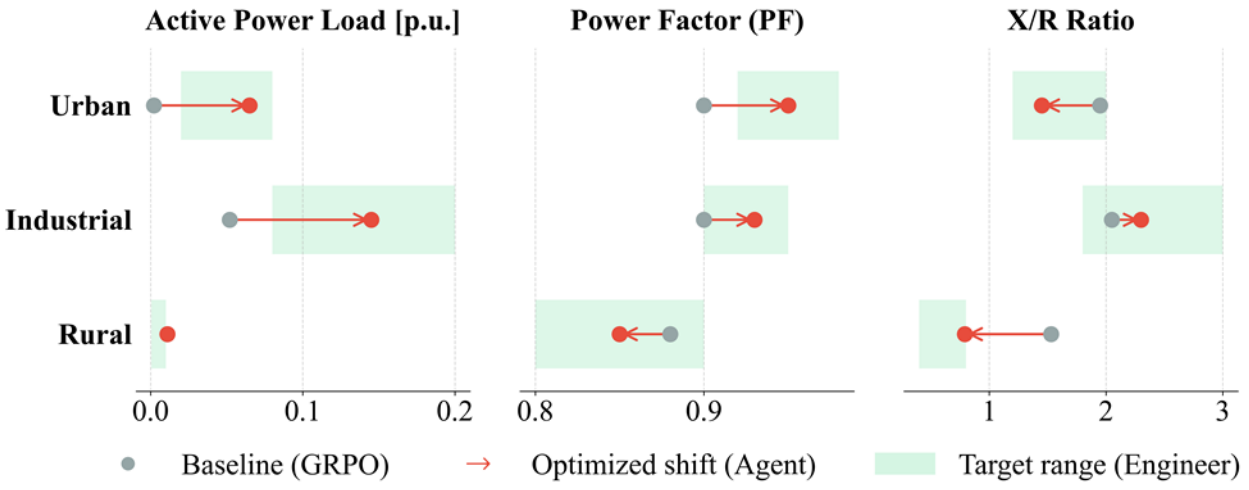

Fig. 8. Parameter shift analysis demonstrating context-aware calibration across different grid scenarios. The grey dots represent the initial parameters generated by the GRPO model. The red arrows illustrate the parameter shifts executed by the refinement agent. The shaded green areas denote engineering reference ranges used for the visualization.

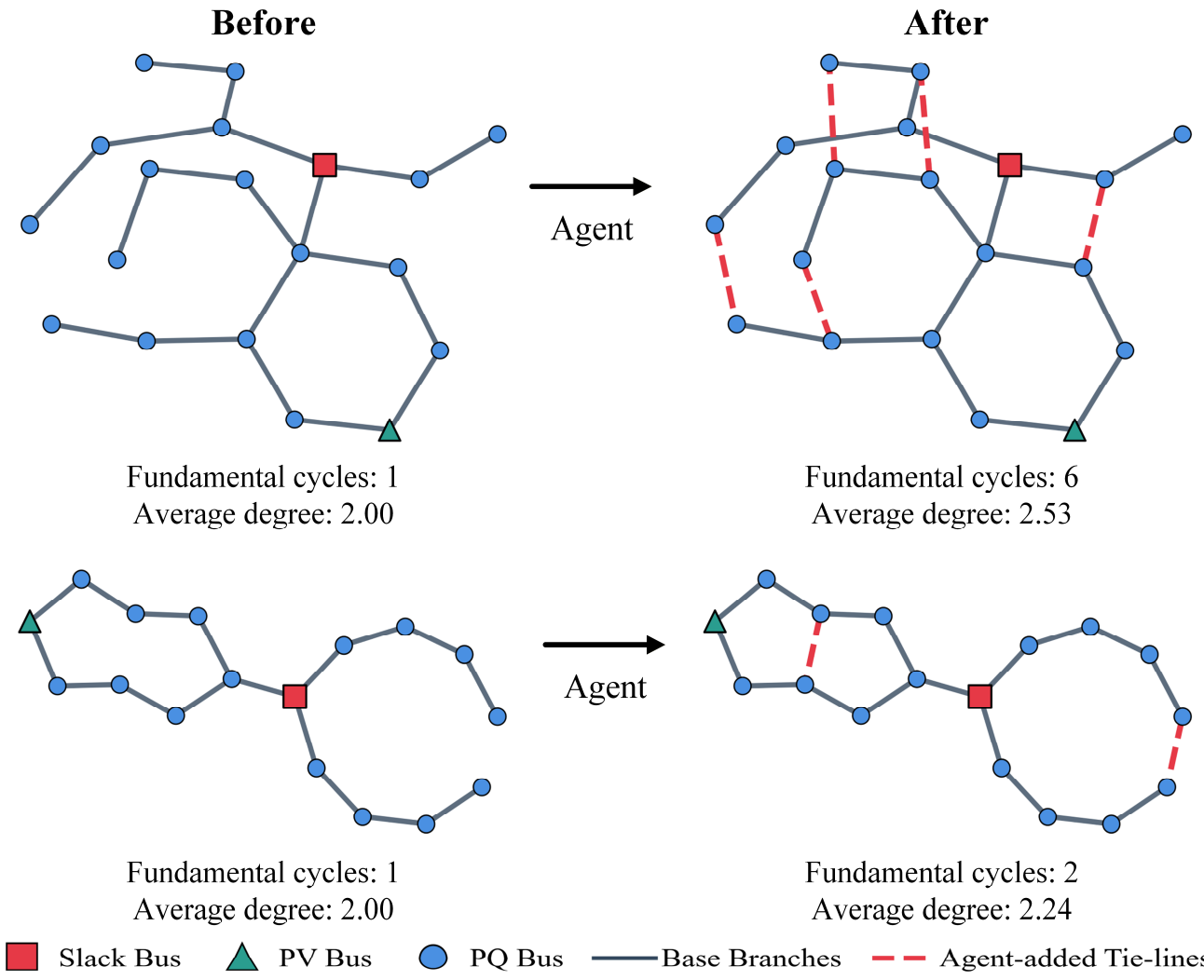

Fig. 9. Topological comparison of feeders before and after agent refinement. The left column displays the base feeders generated by the GRPO model. The right column illustrates the optimized mesh structures. Dashed lines represent the specific tie-lines introduced by the refinement agent. The reported metrics demonstrate the corresponding increase in fundamental cycles and average node degree.

The framework improves engineering value through precise parameter calibration. As illustrated in Fig. 8, a parameter shift analysis demonstrates these specific adjustments across different grid scenarios. The shaded intervals represent engineering reference ranges derived from relevant national standard documents and practical engineering experience. Although the parameters generated by the GRPO model vary, they do not fall precisely within the engineering reference ranges. The optimization agent enhances the practical engineering value of the feeders by shifting the parameters into the appropriate target ranges. For rural grids, the agent maintains low active power loads while reducing the power factor and impedance ratio. For industrial grids, the agent increases the load and impedance ratio to match the physical properties of heavy industrial zones. Urban grid parameters are adjusted to reflect moderate load densities.

In addition to parameter calibration, the agent optimizes network connections. Radial and ring structures are relatively simple and offer limited optimization space. Therefore, topological optimization primarily focuses on mesh grids. The optimization agent evaluates these networks and strategically modifies the edges. It introduces specific tie lines to form closed loops. As illustrated in the topological comparison in Fig. 9, this restructuring increases the number of fundamental cycles and the average node degree. This modification provides alternative power routing paths and transforms the initial feeders into standards-compliant mesh distribution networks.

## D. Suitability for Smart Grid Analysis

Topological connectivity and alternating current power flow convergence are fundamental validation metrics. Application in modern smart grid research requires adherence to complex physical laws under extreme operating conditions. To evaluate physical fidelity and structural robustness, a diverse set of synthesized feeders is subjected to advanced physical stress tests. This set comprises two radial networks (Feeders a and b) and four meshed networks (Feeders c through f). The comprehensive evaluation results are illustrated in Fig. 10 and Table II.

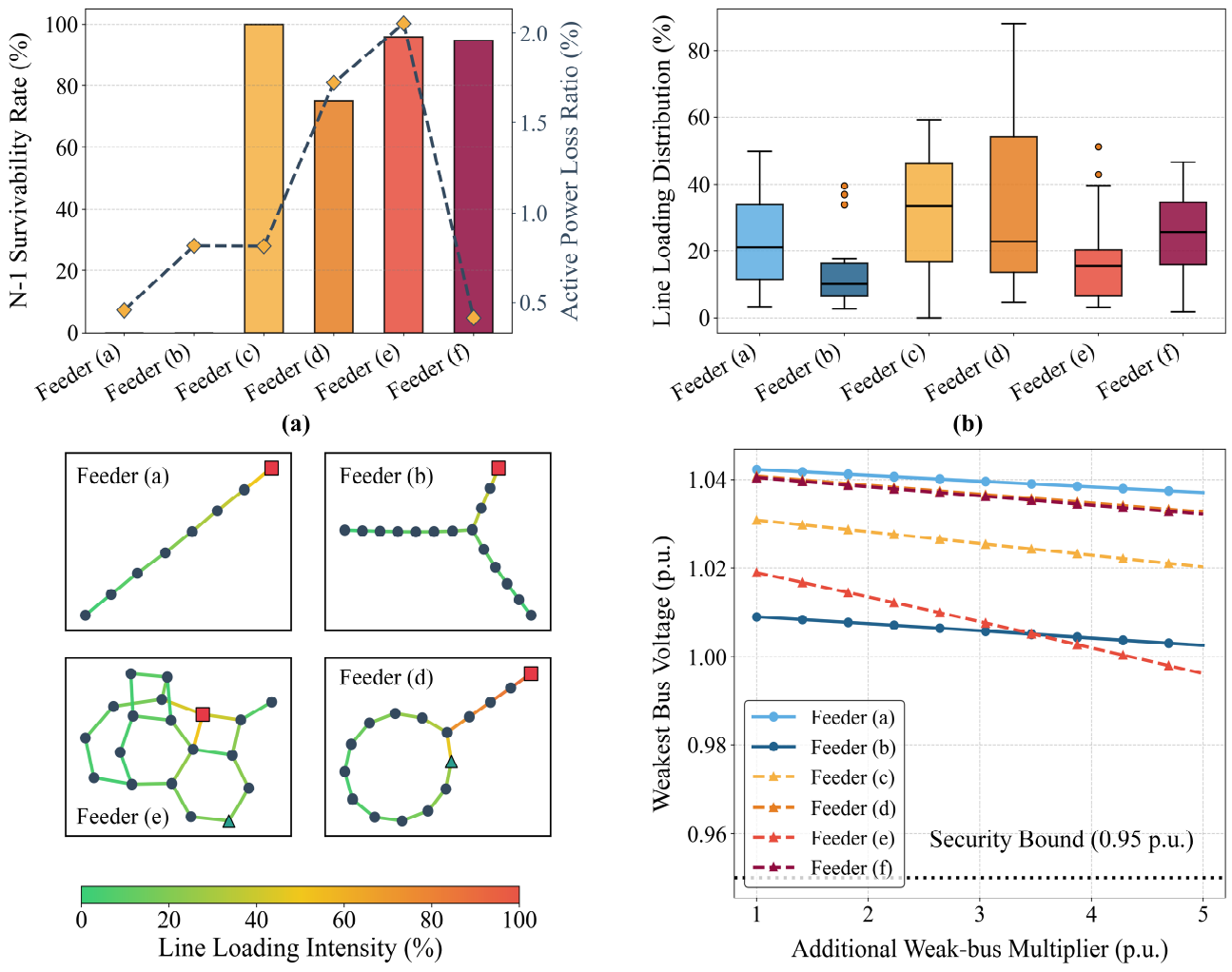

Fig. 10. Physical characteristics of the generated feeders. (a) N-1 survivability rates and active power loss ratios. (b) Line loading distributions. (c) Power flow heatmaps indicating line congestion from low (green) to high (red), where node shapes denote bus types (Slack: square, PV: triangle, PQ: circle). (d) Power-Voltage curves illustrating static voltage stability margins at the weakest bus.

TABLE II
TOTAL GENERATION COST OF GENERATED FEEDERS UNDER ACOPF

| Feeder | (a) | (b) | (c) | (d) | (e) | (f) |
|---|---|---|---|---|---|---|
| Cost ($/h) | 39.19 | 79.74 | 8153.64 | 2253.37 | 3567.30 | 3410.93 |

### 1) Topological Resilience and Flow Distribution

Topological resilience is first assessed via N-1 survivability analysis to evaluate structural robustness (Fig. 10a). Radial feeders inherently demonstrate zero survivability. In contrast, meshed networks exhibit robust security with pass rates ranging from 75% to 100%, while active power loss ratios remain controlled. Beyond structural resilience, line loading intensity serves as a critical indicator of thermal security. As shown in Fig. 10(b), the maximum line loading across all generated networks remains well below the strict 100% physical limit. The flow distribution accurately reflects distinct topological characteristics. In radial configurations (Feeders a and b), the loading peaks at the backbone near the source and decays toward the terminal branches (Fig. 10(c)). In meshed configurations (Feeders c to f), the introduction of tie-lines effectively redistributes the power flow across parallel paths, preventing congestion and balancing network utilization.

### 2) Voltage Stability and Operational Optimization:

Structural robustness is further validated through static voltage stability margins and AC Optimal Power Flow (OPF) evaluations. By tracing power-voltage curves (Fig. 10(d)) under increasing local load multipliers, all feeders exhibit consistent voltage drop. The networks sustain multipliers up to five times the nominal load before breaching the 0.95 per unit security bound. This indicates wide operational margins. Finally, the convergence of the OPF calculations for all tested feeders (Table II) confirms the existence of valid feasible regions. The total generation costs range from $39.19/h to $8153.64/h. This cost variance reflects the operational dynamics under different network scales. The 100% OPF convergence rate observed in the tested cases indicates that the synthetic feeders provide feasible operating points under the specified optimization settings. These operating points satisfy the evaluated nodal voltage limits, branch thermal limits, and generator capacity constraints, suggesting that the generated feeders can support economic dispatch studies.

## E. Ablation Study

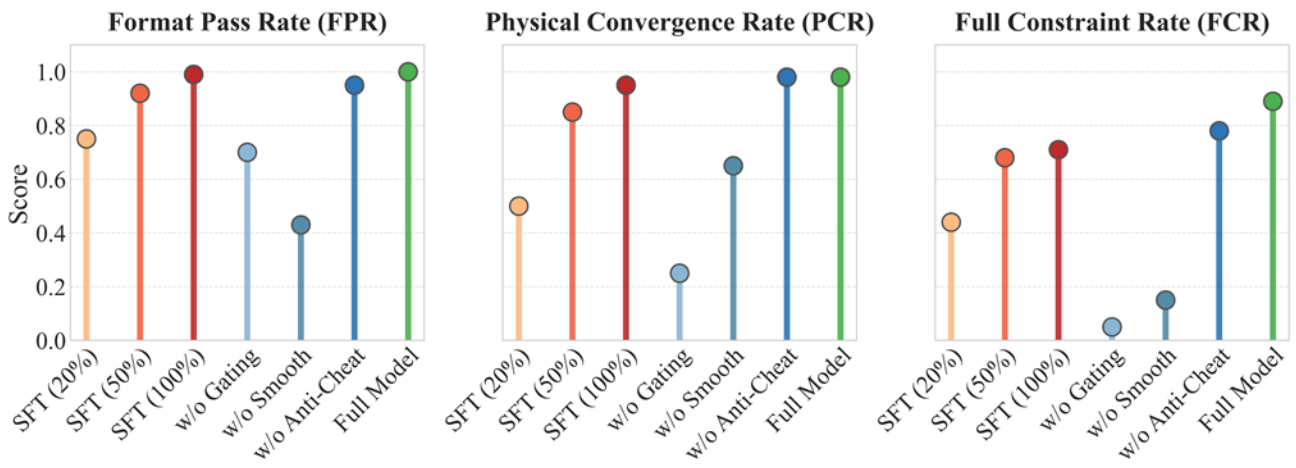

Fig. 11. Ablation study results across different model configurations. The three panels illustrate the format pass rate, physical convergence rate, and full constraint rate. The red series shows the impact of varying the supervised fine tuning dataset size. The blue series demonstrates the performance degradation when specific reward modules are removed. The green series represents the optimal performance of the complete proposed model.

To quantify the impact of dataset size and reward function components, an ablation study is conducted. The results are summarized in Fig. 11. The size of the SFT dataset directly influences the generation quality. This impact is especially pronounced at smaller data volumes. The performance improvements become less significant as the dataset size continues to expand. A restricted dataset provides limited coverage of diverse grid types and topological structures. This lack of diversity hinders the generalization capabilities of the model. A comprehensive dataset is necessary to establish robust structural foundations.

The reward design used during policy optimization has a significant impact on the final generation quality. Removing specific reward constraints exposes the optimization process to reward hacking: once the model identifies loopholes in the reward formulation, it may maximize the reward while violating the intended physical meaning. For example, without the anti-cheat module, the model exploits the network-loss evaluation by driving all nodal loads to zero, thereby obtaining an artificially high score. This behavior produces degenerate grids with no practical engineering value. These results indicate that all designed reward components in the proposed full model are necessary for maintaining physical realism and preventing degenerate solutions.

## IV. Conclusion

This paper has developed an LLM-based framework for synthesizing customized power distribution system feeders from natural language specifications. The main contribution of the framework is to couple data-driven language generation with power-system feasibility checks and engineering refinement, so that the generated feeders are constrained not only by textual requirements but also by physical rules and practical design considerations. In this framework, SFT improves the basic generation capability of the LLM, GRPO aligns the generation policy with engineering preferences and physical constraints, and the agent architecture further optimizes feeder parameters and network topologies. The case studies confirm the effectiveness of the proposed method. Compared with the SFT baseline, SFT+GRPO achieves a format pass rate of 1.000 and a physical convergence rate of 0.980, while increasing the full constraint satisfaction rate from 0.710 to 0.895. The refinement agent also improves the engineering assessment score from 4.19 to 8.05, indicating that the generated feeders better satisfy practical design requirements after calibration and topology refinement. In addition, stress tests and AC optimal power flow analyses show that the synthesized networks maintain reasonable structural and operational performance under the tested conditions. These results suggest that the proposed framework can provide physically feasible and engineering-relevant synthetic feeders for smart grid studies.

Future research will focus on scaling the generative framework to larger distribution networks.